\documentclass[preprint,prd,aps,showpacs,showkeys,nofootinbib]{revtex4}

\usepackage{graphicx}

\usepackage{dcolumn}

\usepackage{bm}
\usepackage{epsfig}

\def\be{\begin{equation}}

\def\ee{\end{equation}}

\begin{document}

\title{Landau-Zener problem in a three-level neutrino system with 
non-linear time dependence}

\author{
P. Ker\"anen $^{a,b}$,
J. Maalampi$^{a,c}$, 
M. Myyryl\"ainen$^{a}$,
and J. Riittinen$^{a}$}

\affiliation{
$^a$ Department of Physics,
P.O.~Box~35, FIN-40014 University of Jyv\"askyl\"a, Finland}
\affiliation{$^b$ Radiation and Nuclear Safety Authority, P.O.~Box~14, FIN-00881 Helsinki}
\affiliation{$^c$ Helsinki Institute of Physics, P.O.~Box~64, FIN-00014 
University of Helsinki}

\date{\today}

\bigskip

\begin{abstract}

We consider the level-crossing problem in a three-level system with 
non-linearly time-varying Hamiltonian (time-dependence $t^{-3}$). We study 
the validity of the so-called independent crossing approximation in the Landau-Zener model
by making comparison with results obtained numerically in density matrix approach.
We also demonstrate the failure of the so-called "nearest zero" approximation of the 
Landau-Zener level-crossing probability integral.
\end{abstract}


\maketitle

\section{Introduction}

The Landau-Zener (LZ) model has been widely used for studying the dynamics of
two-level quantum systems \cite{Landau}. When the Hamiltonian of a time 
dependent system is diagonal, the energy levels  
would in general cross each other. In the presence of non-diagonal terms 
("perturbations") the crossing is, however, avoided  leaving a time-varying 
non-vanishing gap between the levels. If the gap  is sufficiently 
large, transition between the states is strongly suppressed and the system 
behaves adiabatically remaining in its particular energy state. However, 
if the  levels get close to each other the adiabaticity may be violated 
and a transition between the states can take place in the vicinity 
of the points where the diagonal elements of the Hamiltonian (also called diabatic 
energies) cross.

The exact analytical solution of the wave equations for this 
kind of systems can be found only in some special cases. The LZ model 
offers an approximate method for determining  the probability ($P_{\rm LZ}$) 
of the transitions for two-level crossings  by using a quasi-classical approach, 
essentially the WKB approximation.  Originally the LZ-model was applied to 
transitions in diatomic molecules, but it has later found applications also 
in nuclear \cite{LZNucl} and atomic \cite{LZatom} physics, quantum optics 
\cite{LZopt} and laser-driven atoms in time-dependent electric \cite{LZelec} 
and magnetic \cite{LZmagn} fields, as well as in neutrino physics \cite{LZneut,
Wolfenstein}. Recently the LZ-model has been employed also in 
problems related to nanostructures \cite{nano} and Bose-Einstein condensates 
\cite{BEcon}.

In contrast to the two-level case, no general approximate solution, 
i.e. a counterpart of the LZ model, has been found for the level crossing 
problem of multilevel systems with three or more states. That is, there 
is no general rule which would estimately tell the state of the multilevel 
system after the system has passed the level-crossing region when the state of the
system before entering the region is given and 
which would take into account the effect of level crossing. 
Such theoretically derived rules 
exist only for some special cases (see e.g.\cite{Sinitsyn} 
and references therein). 

There are also some results which lack mathematical 
justification but are verified 
with a high accuracy by numerical studies \cite{Smatrix}. One such result 
concerns with a three-level 
system, where diabatic energies  behave all differently as a function of 
time so that energy levels cross with one another (see Fig.~1). It has 
been shown numerically that the probability for transitions in this kind 
of systems can be obtained  as a product of the relevant two-state LZ 
probabilities given that the Hamiltonian describing the system has a 
{\it linear time dependence}. This is what one would expect in the case 
when the crossing points of  diabatic energy levels are well 
separated, since then only two  states at a time are close to each 
other while the third one is separated from these two with a larger energy gap. 
What is more surprising is that this simple rule is shown to be valid even 
when all the crossing points  are close to each other \cite{Smatrix}. 
That is, the so-called independent crossing approximation works not only 
in the case where the crossings are well separated but also in the 
case where they overlap.

In this paper we will study this kind of three-level system 
in the case of a {\it nonlinear time dependence}. We have come across 
with such a situation when  studying evolution of neutrino states in 
environments where the time dependence of the Hamiltonian is due to a 
varying density of the background medium \cite{KMMR}. The matter affects different 
neutrino types differently, which results in level crossing 
phenomena. These effects depend on the density distribution of 
the matter, and they are typically non-linear. We will consider $r^{-3}$ dependence, 
which is a typical situation in supernovae. This implies the time dependence 
$t^{-3}$ as light neutrinos are highly relativistic. We will 
solve the evolution of the three-level system numerically by using the 
density matrix approach, which we suppose to give the correct description 
of the behavior of the system. We then make a comparison of the 
results of the numerical study with the results obtained by the LZ theory 
and the independent crossing approximation.

\section{Three-level neutrino system}

We will consider a system of the electron neutrino ($\nu_e$), another active 
neutrino ($\nu_{a}, a=\mu\,{\rm or}\, \tau$), and a so-called sterile 
neutrino ($\nu_s$). These neutrinos behave differently 
in a medium: $\nu_e$ encounters both charged and neutral current weak 
interactions, $\nu_{a}$ neutral current interactions only, and $\nu_s$ 
has no interactions with matter at all. The interactions with a 
medium appears as a "potential energy" ($V_e, V_{a}, V_s$), which is 
proportional to the density of the matter $\rho$ and the coupling constant 
of the weak force $G_F$ \cite{Wolfenstein}. The density $\rho$ 
varies as a function of the location of the   neutrino, or the 
distance $r$ the neutrino has traveled from its creation point.

\begin{figure}[h!]
\centering
\epsfig{file=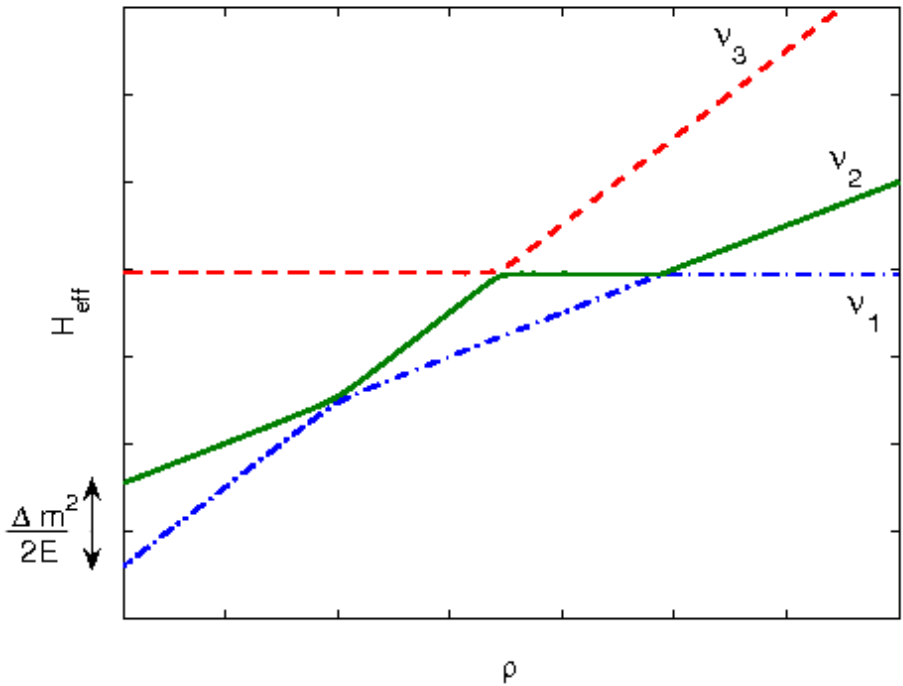,height=10cm}
\caption{{\small Diabatic energy levels of a system consisting of the electron neutrino 
(the steepest straight line), a sterile neutrino (the next steepest line) 
and the muon or tau neutrino (the horizontal line) in a medium of varying 
density with normalization defined in the text. Non-diagonal elements of 
the Hamiltonian makes the system to avoid the crossings making the adiabatic 
states evolve as indicated by the different line formats in the figure.}}
\label{fig1}
\end{figure}

The evolution of the system is described by the Schr\"odinger equation 
(the Hamiltonian is assumed real 
for simplicity)
\begin{equation}
i\left(
\begin{array}{c}
\dot{\nu_e} \\
\dot{\nu_s} \\
\dot{\nu_a} \\
\end{array}
\right)
=\left( \begin{array}{ccc}
H_{ee} + V_{e}& H_{es} & H_{ea} \\
H_{es} & H_{ss} +V_s & H_{sa} \\
H_{ea} & H_{sa} & H_{aa}+V_a \\ 
\end{array} \right)
\left(
\begin{array}{c}
  \nu_e \\
  \nu_s \\
  \nu_a \\
\end{array}
\right). 
\label{genU} 
\end{equation}
The elements $H_{bc}$, which constitute the time-independent part of the 
Hamiltonian, correspond to the Hamiltonian in vacuum, the non-diagonal 
elements ($b\neq c$) arising purely from the mixing of neutrino 
flavours. The neutrino mixing has been observed in experiments through 
neutrino oscillation phenomena among the neutrinos created in the 
atmosphere by cosmic rays and among neutrinos originating in the 
nuclear processes taking place in the core of the Sun. In the following 
we will not fix our consideration to any particular physical 
situation but analyze the three-level problem generically.

The $H_{bc}$'s are the elements of the Hamiltonian in vacuum presented 
in the $(\nu_e,\nu_s,\nu_a)$ 
flavor basis.  They are obtained from
\begin{equation}
U\frac{1}{2E}\left( \begin{array}{ccc}
m_1^2& 0 & 0 \\
0 & m_2^2 & 0 \\
  0 & 0 & m_3^2 \\ 
\end{array} \right)U^{\dag},
\label{UdagU}
\end{equation}
where the matrix $U$ can be parameterized in terms of three mixing 
angles $\theta_i, i=1,2,3$ as follows:
\begin{equation}
U=\left( \begin{array}{ccc}
c_{1}c_{3} & s_{1}c_{3} & s_{3} \\
-s_{1}c_{2}-c_{1}s_{2}s_{3} & c_{1}c_{2}-s_{1}s_{2}s_{3} & s_{2}c_{3} \\
s_{1}s_{2}-c_{1}c_{2}s_{3} & -c_{1}s_{2}-s_{1}c_{2}s_{3} & c_{2}c_{3} \\
\end{array} \right),
\label{Umatrix}
\end{equation}
where $c_{i} = \cos{\theta_{i}}$ and
$s_{i} = \sin{\theta_{i}}$. We will consider the mixing angles and neutrino energy as free parameters and we fix their values as is suitable for our purposes. 

The mass terms in the diagonal elements of the Hamiltonian result from the 
relativistic approximation $E_i=\sqrt{p^2+m_i^2}\simeq p+m_i^2/2E$ once the 
irrelevant overall phase $\exp 
(ip)$ is omitted. In other words, the behavior of the states in the resonances 
depends only on the energy differences between the states, not on the absolute 
values of energies. Similarly, one can also normalize the potential energies. 
As is common, we will take out the neutral current contribution in a form of 
an overall phase, leaving the following effective matter potentials for the 
electron neutrino, the other active neutrinos, and the 
sterile neutrino, respectively \cite{Wolfenstein}:
\begin{eqnarray}
V_e & = & \sqrt{2}G_F N_e \nonumber \\
V_a & = & 0 \\
V_s & = & \frac{1}{\sqrt{2}}G_F N_e. \nonumber 
\end{eqnarray}
We have assumed ordinary matter where $N_e \propto \rho$, and we have 
for simplicity assumed that $N_n = N_e$.

The starting point of the LZ approximation is an analytic continuation 
of the classical action integrals into the complex coordinate plane. 
In our case this results in the 
following transition probability between the two states $\nu_b$ and $\nu_c$ 
($b,c=e,s,a$) (in natural units $\hbar=c=1$) \cite{Landau}
\be
P^{LZ}_{bc}=\exp\left[{{-2{\rm Im}\int^{t_0}\ {\rm d}t\sqrt{((H_{bb}+V_b)-(H_{cc}+V_c))^2
+4H_{bc}^2}}}\right].
\label{PLZ}
\ee
Here $t_0$ is a point (a branch point) on the upper half-plane of the 
complex time, the one nearest to the real time axis,  where the quantity 
under the square root vanishes. The integrand is the energy gap 
between the two eigenstates (quasi-energies) of the $2\times 2$ sub-Hamiltonian spanned 
by the states $\nu_b$ and $\nu_c$. 
Of course, the eigenstates of the submatrices generally differ from  
the proper energy eigenstates of the full $3\times 3$ Hamiltonian.

In the case where the time dependence is linear the integral in (\ref{PLZ}) 
can be easily solved and the probability is given by the following  
simple expression \cite{Landau}:
\begin{equation}
P^{LZ}_{bc}=\exp[{-{\pi \gamma_{bc}}/{2}}],
\label{PLZlin}
\end{equation}
where the parameter $\gamma_{bc}$ is proportional to the square of the 
perturbation connecting the states $b$ and $c$, i.e. of $H_{bc}$, and 
inversely proportional to the slope difference of the unperturbed energy 
levels at the crossing point $r_R$:
\begin{equation}
\gamma_{bc}=\frac{4 H_{bc}^2}{({d}/{dr})(V_b-V_c)\vert_{r=r_R}}.
\label{eq:gamfac}
\end{equation}
In contrast to the linear case, in non-linear cases the integral (\ref{PLZ}) 
cannot in general be presented in closed form  but one has to evaluate 
it numerically or approximate it with a suitable series expansion \cite{Pantaleone}. 
We will evaluate the integral numerically.

\section{Results}

We now describe the results of our  study of the validity of the independent 
crossing approximation in a three-neutrino case with a non-linear time 
dependence of energy levels. 

We first compute numerically the time evolution of neutrino states using 
the density matrix formalism. That is, we solve the matrix equation
\be
\dot{\rho} = i \left[ \rho,H \right],
\ee
where $\rho$ is the density matrix. The diagonal elements 
$\rho_{ee},\:\rho_{ss}$ and $\rho_{aa}$ of the density matrix 
give the probability for the system with a given initial state 
at high positive density to end into the states 
$\nu_e$, $\nu_s$ and $\nu_a$, respectively, at high negative density. Negative density 
is, of course, just a formal notion; physically, a neutrino at negative 
density is interpreted as an antineutrino at positive density as the matter potential
$V$ has an opposite sign for neutrino and its antiparticle.
(In the cases we will consider it will actually be enough to compute the evolution 
to zero density  as  no level crossing will take place in 
negative densities). The relevant contribution to the level crossings 
is achieved in a limited range where the  matter potentials $V_b$ ($b=e,s,a$) 
are of the order of the perturbations $H_{bc}$. The quantity that 
characterizes the size of this range is the transition width $\delta r_{bc}$, 
which we define as follows:
\be
\delta r_{bc}={\Bigg \vert}\frac{ H_{bc}}{(d/dr)(V_b-V_c)\vert_{r=r_R}}{\Bigg \vert}.
\label{width}
\ee
When one is well outside this range practically no level crossing takes place 
but the system remains in the adiabatic energy state. 

The ($\nu_e,\nu_s, \nu_a$) content of the eigenstates of the Hamiltonian 
of course varies due to the mixing of the neutrinos. The varying
ceases when the potential energy terms  $V_b$ overrule the other 
terms $H_{bc}$ of the Hamiltonian and the system ends to one of the states 
$\nu_e$, $\nu_s$ and $\nu_a$ at large negative densities. As no level 
crossing takes place at negative densities in our case, the flavour 
states at large negative densities have one-to-one correspondence with 
mass eigenstates at zero density.

By adjusting the values of the parameters of the Hamiltonian one can 
move the locations of the crossing points in respect to each other. 
We have used this in order to test the dependence of the independent crossing 
approximation  on the distance of the crossing points. As far as the 
crossing points are well separated, that is, their transition width 
ranges are far from overlapping, the crossings can be clearly considered 
as independent two-level systems covered by the Landau-Zener model. 
Then the probability for the level crossing is given by 
the product of the Landau-Zener probabilities of the two-level crossings on 
the route. 

When the crossing points are so close to each other that their 
corresponding widths overlap, the situation is different and it is 
not a priori justified to consider the  two-level transitions as independent.  
Nevertheless, numerical studies show that in the case of 
linear time-dependence, the independent crossing 
approximation still gives the correct result \cite{Smatrix}. We have verified 
the validity of this result for three-neutrino system in the case of linearly varying 
matter density. In the following we will study the non-linear case.

We assume that our system consists in the beginning at high matter density 
from the adiabatic state $\nu_3=\nu_e$ purely, and we calculate the probability 
for having the system in the adiabatic state 
$\nu_1$ in the end at zero density (corresponds to the state $\overline\nu_e$ at 
large negative density).  As one can deduce from Fig. 1, the transition 
from the adiabatic state $\nu_3$ to the adiabatic state $\nu_1$ 
requires non-adiabatic behaviour of the system in 
two crossings it encounters, i.e. in the 21- and 32-crossings. In the LZ model 
the probability 
$P^{LZ}_{31}$ is thus given by
\be
P^{LZ}_{31}=P^{LZ}_{32}P^{LZ}_{21}.
\label{P_LZ}
\ee

In Fig.~2 we present the probability of the $3\to 1$ transition calculated 
by the LZ approximation ($P^{LZ}_{31}$) and numerically using the density 
matrix formalism ($P^{\rho}_{31}$) for different distances between the 
two crossing points. We have changed the locations of the crossing points 
by varying the parameter 
$\Delta m^2$, which is the difference between the adiabatic energy states 
$\nu_1$ and $\nu_2$ at zero density (see Fig.~1). 
At high values of $\Delta m^2$ the crossing points are well separated 
and their corresponding transition regions do not overlap. With decreasing 
$\Delta m^2$ the crossing points come closer to each other, and at 
$\Delta m^2=1\times 10^{-6} eV^2$  the two transition regions start to overlap. At 
$\Delta m^2=0.5\times 10^{-6} eV^2$ the regions are equally wide and overlap 
completely. At $\Delta m^2=0.2\times 10^{-6} eV^2$ they are again fully separated.

\begin{figure}[h!]
\centering
\epsfig{file=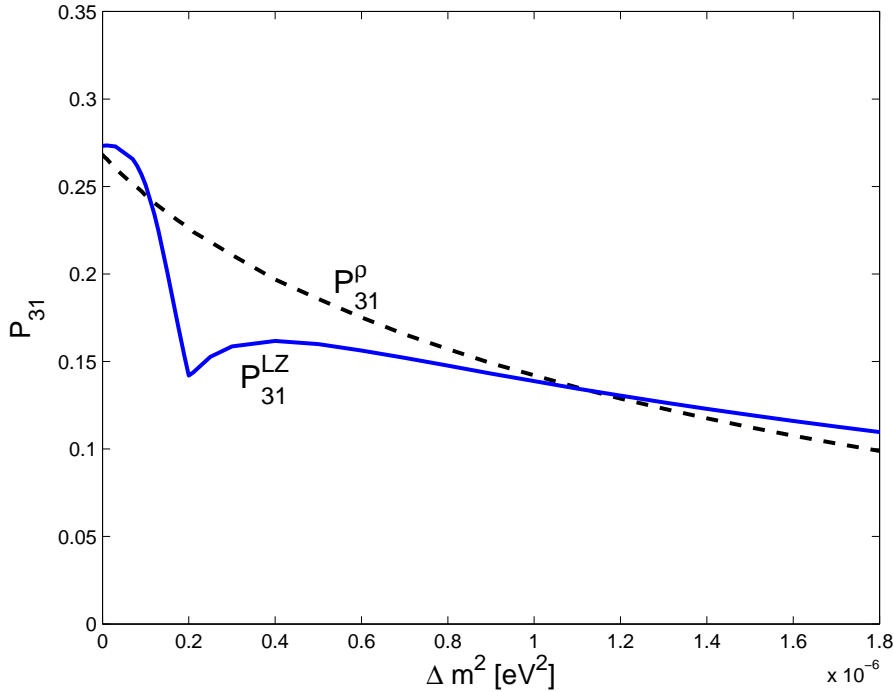,height=10cm}
\caption{\small The probability of the $3\to 1$ transition as a function of the parameter
$\Delta m^2$ as calculated by using the LZ 
theory and the independent crossing approximation (solid line) and numerically 
using the density matrix formalism (dashed line). The two transition regions overlap
fully at $\Delta m^2=0.5\times 10^{-6} eV^2$.}
\label{fig2}
\end{figure}

As the plot in Fig.~2 shows, the Landau-Zener model seems to work in  
our non-linear case relatively well, though underestimating the transition 
probability to some extent.  Anyway, the overlapping of
transition regions seems not to dramatically affect the validity of the
independent crossing approximation{\footnote[1]{\footnotesize The LZ probability (\ref{PLZ}) 
needs to be modified in the extreme inadiabatic limit. We have done it 
according to the ansatz of ref. \cite{Pantaleone}}}. 
The situation is actually  better 
than the Fig.~2 indicates, as we will now explain.

Fig.~2 shows a curious 
behaviour of $P^{LZ}_{31}$ when $\Delta m^2$ is  decreased beyond 
the maximal overlap region. The LZ result seems to be in a serious contradiction 
with the numerical result of the density matrix approach 
around the point $\Delta m^2=0.2\times 10^{-6} eV^2$. This 
behaviour is actually an artifact of  simplistic use of the LZ theory
and can be understood as follows. As mentioned, in the integral (\ref{PLZ}) 
the  upper bound of the integration $t_0$ is a point on the upper half-plane of 
the complex time nearest to the real time axis, where the quasi-energy gap between 
the two states vanishes. In general, the closest branch point gives the 
dominant contributions and the contributions of the other branch points 
with a larger imaginary part are exponentially small, as assumed in the LZ theory.
 
In our case it 
happens, however, that at $\Delta m^2=0.2\times 10^{-6} eV^2$ there are two branch 
points ($(t_0)_1$ and $(t_0)_2$) with an equal imaginary part, and actually the role of 
the closest branch point moves from one $t_0$ 
to the other at this value of $\Delta m^2$. That is, different 
$t_0$ gives the dominant contribution for large values of $\Delta m^2$ 
than for small values. This is illustrated in Fig.~3, where we plot 
$P^{LZ}_{31}$ for both of these two zeros. Obviously, the 
contributions of the both zeros (actually, of all zeros) should be 
taken into account when evaluating $P^{LZ}_{31}$. It goes beyond 
the scope of the present work to find out an appropriate way to combine the 
different contributions coherently (for a discussion on this matter 
for a two-level systems, see \cite{KalleAntti}). Nevertheless, it is quite 
obvious that once this 
summation is properly done  the odd 
behaviour of the $P^{LZ}_{31}$ in Fig.~2 will level off and that the the 
result obtained by the LZ theory 
and independent crossing approximation gets closer to the numerical result obtained
using the density matrix approach than the Fig.~2 shows.

\begin{figure}[h!]
\centering
\epsfig{file=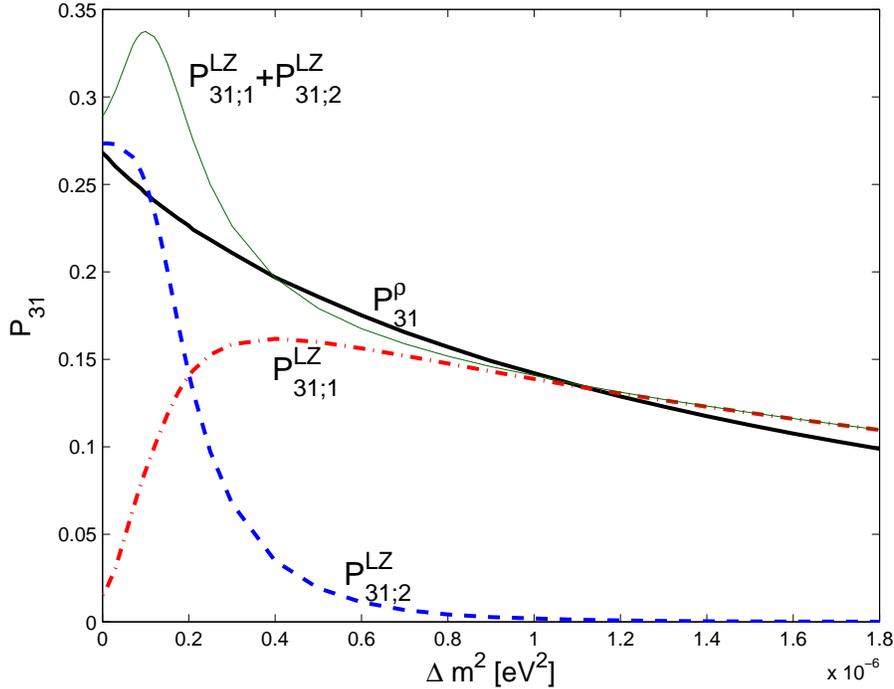,height=10cm}
\caption{\small The contributions corresponding to the two dominating branch 
points $(t_0)_1$ and 
$(t_0)_2$ to the transition 
probability $P^{LZ}_{31}$ (dash-dotted and dashed lines, respectively). 
Also shown are the numerical result obtained by the density matrix 
approach (thick solid line) and the
incoherent sum of the contributions from the two branch points (thin solid line).}
\label{fig3}
\end{figure}

\section{Conclusions}

We have studied  in this paper the validity of the independent crossing 
approximation  for a  three-level system with non-linear time dependence 
in the framework of the Landau-Zener model. In particular, we have investigated
the case where the transition regions of two level crossings overlap, and we have
found  that considering the crossings as separate still gives a reasonably good 
approximation for transition probability as compared with the numerical 
estimate obtained by using
density matrix approach. We have demonstrated also that the simplistic 
use of the Landau-Zener  theory, where the tunneling integral is 
calculated only for the "nearest zero" in the complex time plane, leads
in some occasions to very inaccurate results.

\begin{acknowledgments}

We are grateful to Kimmo Kainulainen for many useful discussions. 
This work has been supported by the
Academy of Finland under the contracts no.\  104915 and 107293.

\end{acknowledgments}


\begin{thebibliography}{99}

\bibitem{Landau} L. Landau, Phys. Z. Sowj. 2 (1932) 46; C. Zener, Proc. R. Soc. A 137 (1932) 696;
E.C.G. St\"uckelberg, Helv. Phys. Acta 5 (1932) 369; E. Majorana, Nuovo Cimento 9 (1932) 43. For
a pedagogical presentation, see L. D. Landau and E. M. Lifshitz, Quantum mechanics : non-relativistic theory. 

\bibitem{LZNucl}  Y. Abe and J.Y. Park, Phys. Rev. C 28 (1983) 1355.

\bibitem{LZatom} A.P. Baede, {\it Advances in Chemical Physics}, vol. 30, eds. I. Prigogine and S.A. Rice (Wiley, London, 1975) p.463; S.D. Drell, N.M. Kroll, M.T. Mueller, S.J. Parke and M.A. Ruderman, Phys. Rev. Lett. 50 (1983) 644.

\bibitem{LZopt} C.E. Carrol and F.T. Hioe, J. Phys. A: Math. Gen. 19 (1986) 1151.

\bibitem{LZelec} D.A. Harmin, Phys. Rev. A 44 (1991) 433.

\bibitem{LZmagn} A. Messiah, {\it Quantum Mechanics} vol. II (North Holland, Amsterdam, 1975).
\bibitem{LZneut} S.P. Mikheyev and A. Yu. Smirnov, Sov. J. Nucl. Phys. 42 (1985) 913; H.A. Bethe, Phys. Rev. Lett. 56 (1986) 1305; C.W. Kim, W.K. Sze and Shmuel Nussinov, Phys. Rev. D 35 (1987) 4014, and Phys. Lett. B 184 (1987) 403. 

\bibitem{Wolfenstein} L. Wolfenstein, Phys. Rev. D 17 (1978) 2369.
\bibitem{nano} D.V. Averin, Phys. Rev. Lett. 82 (1999) 3685; A.D. Armour and A. MacKinnon, Phys. Rev. B 66 (2002) 035333.

\bibitem{BEcon} Q. Niu, X.-G. Zhao, G.A. Georgakis, and M.G. Raizen, Phys. Rev. Lett. 76 (1996) 4504; V.A. Yurovsky and A. Ben-Reuven, Phys. Rev. A 63 (2001) 043404.

\bibitem{Sinitsyn} N. Sinitsyn, Thesis, Texas A\& M University (May 2004) and references therein.

\bibitem{Smatrix} S. Brundobler and V. Elser, J. Phys. A: Math. Gen. 26 (1993) 1211.
\bibitem{KMMR} P. Ker\"anen, J. Maalampi, M. Myyryl\"ainen, and J. Riittinen, Phys.
Lett. B 597 (2004) 374, and work in progress.
\bibitem{Pantaleone} T.K. Kuo and J. Pantaleone, Phys. Rev. D 39 (1989) 1930.
\bibitem{KalleAntti} J.P. Davis and P. Pechukas, J. Chem. Phys. 64 (1976) 3129; K.-A. Suominen and B.M. Garraway, Phys. Rev. A 45 (1992) 374; N.V. Vitanov and K.-A. Suominen, Phys Rev. A 59 (1999) 4580; N.V. Vitanov, Phys. Rev. A 59 (1999) 988.



\end{thebibliography}
\end{document}